\documentclass[twoside]{article}
\usepackage{epsfig.sty}
\newcommand{\beq}{\begin{eqnarray}}
\newcommand{\eeq}{\end{eqnarray}}
\begin{document}
\title{Neutrino Excess and Dark Photons}
\author{Leonard S. Kisslinger\\
Department of Physics, Carnegie Mellon University, Pittsburgh, PA 15213}
\date{}
\maketitle
\begin{abstract}
  Recently an excess of electron neutrino, $\nu_e$, events in a MiniBooNE
  experiment has been reported. An attempt to explain this anamoly was made
  in a theoretical analysis by introducing a new light Higgs Boson. In the
  present work we use a new neutrino interaction Z bosons producing Dark
  Photons to estimate an excess of $\nu_\mu \rightarrow \nu_e$ events.
\end{abstract}
\noindent
PACS Indices:14.60.Lm,13.15.+g,14.60.Pq
\vspace{1mm}

\section{Introduction}

Recently an excess of electron neutrinos in $\nu_\mu \rightarrow \nu_e$ events
has been found\cite{mini18}. In an attempt to explain this anomaly a 
theoretical analysis was carried out with a light Higgs Boson\cite{asa18}.

Cosmic Microwave Background Radiation experiments have estimated Dark Matter
density to be  about 23 \% of the Universe\cite{wmap13}. Sterile neutrinos are
a well-known source of Dark Matter. The MiniBooNE Collaboration\cite{mini13}
estimated the mass difference between a sterile $\nu_4$ and  a standard
neutrino $\nu_1$ as $ m_4^2-m_1^4 \simeq 0.06 (eV)^2$. Another possible Dark
Matter particle is a Dark photon a vector boson (quantum spin=1), while a
neutrino is fermion (quantum spin=1/2, so Dark Photons would be a very
different kind of Dark Matter particle than a sterile neutrino.

There is a great interest in Dark Photons.  Prospects for a direct search for
Dark Photons at the Fermilab are discussed in a recent review\cite{mxl17},
which also gives parameters needed in the present work.

\section{$\nu_\mu \rightarrow \nu_e$ oscillation excess with two
  sterile neutrinos and Dark Photons}

The probability  of $\mu$ to $e$ neutrino oscillation with two sterile
neutrinos was estimated\cite{lsk16}.

The transition probability for a muon neutrino to oscillate to an electron
neutrino, $\mathcal{P}(\nu_\mu\rightarrow \nu_e)$, was obtained from a 5x5
U matrix and the neutrino mass differences
$\delta m_{ij}^2=m_i^2-m_j^2$ for a neutrino beam with energy $E$ and baseline
$L$ by\cite{lsk16}
\beq
\label{Pue-1}
 \mathcal{P}(\nu_\mu \rightarrow\nu_e) &=& Re[\sum_{i=1}^{5}\sum_{j=1}^{5}
U_{1i}U^*_{1j}U^*_{2i}U_{2j} e^{-i(\delta m_{ij}^2/E)L}]\simeq 0.1 \; .
\eeq
\newpage

In addition to  $\mathcal{P}(\nu_\mu\rightarrow \nu_e)$ being estimated with
two sterile neutrinos as shown in Eq(\ref{Pue-1}) we are considering an
additional term with a Dark Photon, illustrated in Figure 1, with $A$ a
Dark Photon.

\begin{figure}[ht]
\begin{center}
\epsfig{file=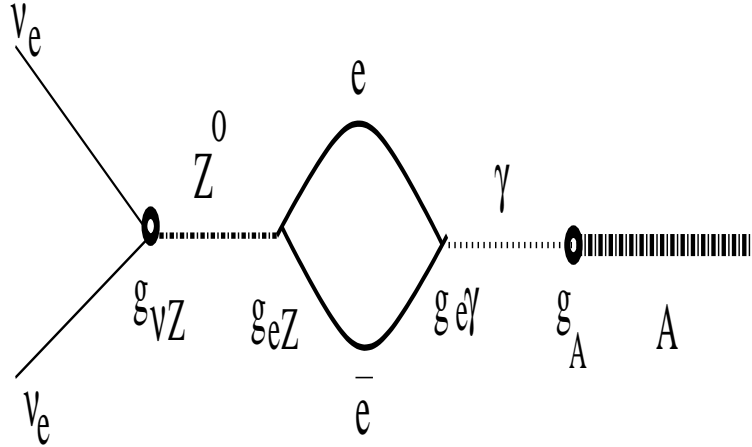,height=6cm,width=10cm}
\caption{Dark Photon electron neutrino interaction}
\label{Figure 1}
\end{center}
\end{figure}

From Ref\cite{mxl17} $M_A$ = 1MeV--10GeV in the di-electron channel.
We estimate the coupling constant $g_A$ from Refs\cite{mxl17,tdl19}.
The coupling constants $g_{\nu Z}$ and $g_{eZ}$ are the weak couplings of the
Z boson, and $g_{e \gamma }$ is the standard electromagnetic coupling of a
photon to an electrically charged particle like the electron.
From Ref\cite{tdl19}  $g_A\simeq 10^{-4}$ to $10^{-3}$.

The $Z-\gamma$-electron correlator is illustrated in Figure 2
\begin{figure}[ht]
\begin{center}
\epsfig{file=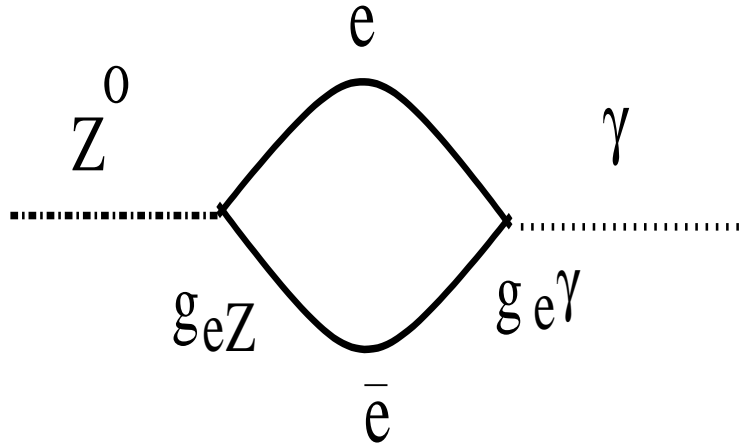,height=6cm,width=10cm}
\caption{$Z-\gamma$-electron correlator}
\label{Figure 2}
\end{center}
\end{figure}

\newpage

The electron correlator in momentum space, corresponding to Figure 2, is
\beq
\label{Hcorrelator}
  \Pi_{ee}^{\mu \mu }(p)& =& g_{eZ}g_{e \gamma}  \int \frac{d^4 k}{(2 \pi)^4} 
Tr[S(k) \gamma^\mu S(p-k)\gamma^{mu}] \nonumber  \\
          S(k) &=& \frac{\not\!k + M_t}{k^2-M_t^2} \nonumber \\
Tr[S(k) \gamma\mu S(p-k)\gamma^{\mu}]&=& \frac{M_e^2 Tr[\gamma^\mu
\gamma^\mu]-k_\alpha (p-k)_\beta Tr[\gamma^\alpha \gamma^\mu \gamma^\beta 
\gamma^\mu]}{(k^2-M_e^2)[(p-k)^2-M_e^2]} \nonumber \\
&=& \frac{4[M_e^2+k\cdot(p-k)}{(k^2-M_e^2)[(p-k)^2-M_e^2]} \; .
\eeq
Thus, with the notation $ \Pi_{ee}^{\mu \mu }(p)= \Pi_{ee}(p)$
\beq
\label{Hcorrelator2}
  \Pi_{ee}(p)& =&4 g_{eZ}g_{e \gamma}  \int \frac{d^4 k}{(2 \pi)^4}
\frac{4M_e^2+k\cdot(p)-k^2}{(k^2-M_e^2)[(p-k)^2-M_e^2]} \; .
\eeq

Using
\beq
\label{Io}
\int \frac{d^4 k}{(2 \pi)^4}\frac{1}{(k^2-M_e^2)[(p-k)^2-M_e^2]}&=&
\frac{2M_e^2-p^2/2}{(4 \pi)^2} I_0(p) \nonumber \\
 I_0(p)= \int_0^1 \frac{d\alpha}{\alpha(1-\alpha)p^2 -M_e^2} \; . 
\eeq

 From Eqs(\ref{Hcorrelator},\ref{Hcorrelator2},\ref{Io})
\beq
\label{correlator-3}
\Pi_{ee}(p)& =& \frac{4 g_{eZ} g_{e \gamma}}{3(4 \pi)^2}[(M_e^2-\frac{7}{4}p^2)
I_0(p)+\frac{21}{2} p^2-2M_e^2] \; .
\eeq

Using $g_{eZ}\simeq 0.53 \simeq g_{e \gamma}$
\beq
\label{Hcorrelator-4}
\Pi_{ee}(p)& \simeq& 2.37 10^{-3} [(M_e^2-\frac{7}{4}p^2)I_0(p)+
\frac{21}{2} p^2-2M_e^2] \; .
\eeq

From $\Pi_{ee}(p)$ in Eq(\ref{Hcorrelator-4}) and the modification of
$\mathcal{P}(\nu_\mu \rightarrow\nu_e)$=$\Delta \mathcal{P}(\nu_\mu
\rightarrow\nu_e)=g_{\nu Z}g_A\Pi_{ee}(p)\simeq 3.73 10^{-5}\Pi_{ee}(p)$,
with $g_{\nu Z} \simeq 0.0373$, $g_A\simeq 10^{-3}$,
we estimate $\Delta \mathcal{P}(\nu_\mu \rightarrow\nu_e)$
\beq
\label{DelPue}
\Delta \mathcal{P}(\nu_\mu\rightarrow\nu_e &\simeq& 8.84 10^{-8}
[(M_e^2-\frac{7}{4}p^2) I_0(p)+\frac{21}{2} p^2-2M_e^2] \; .
\eeq

\newpage

\section{Conclusions}
From the results for the modification of the proability of $\mu$ to $e$
neutrino oscillation, $\Delta \mathcal{P}(\nu_\mu \rightarrow\nu_e)$, by the
emmision of a Dark Photon vis Z-boson emission is shown in Figure 1, from
Eq(\ref{DelPue}) we conclude that the excess electron neutrinos
found be the MiniBooNE Collaboration\cite{mini18} cannot be explained by the
emission of Dark Photons.
\vspace{5mm}

\Large
{\bf Acknowledgements}
\vspace{3mm}

\normalsize
This work was carried out while LSK was a visitor at Los Alamos
National Laboratory, Group P25. The author thanks Dr. Ming X. Liu for
help in choosing the parameters and Dr. William Louis for information about
recent neutrino oscillation exleriments


\vspace{5mm}

\begin{thebibliography}{99}
\bibitem{mini18} A.A Aguilar et. al. (MiniBooNE Collaboration), Phys. Rev. Lett.
  {\bf 121}, 221801 (2018)
\bibitem{asa18}J. Asaadi, E. Church, R. Guenette, B.J.P. Jones and A.M. Szelc,
  Phys. Rev. D {\bf 97}, 075021 (2018)
\bibitem{wmap13} G. Hinshaw et. al. (WMAP), arXiv:1212.5226[astro-ph] (2013)
\bibitem{mini13} A.A. Aguilar-Arevalo $et.\;al.$ (MiniBooNE Collaboration),
  Phys. Rev. Lett. {\bf 110}, 161801 (2013)
\bibitem{mxl17} Ming Xiong liu, Mod. Phys. Lett. A {\bf 32}, 1730008 (2017) 
\bibitem{lsk16} Leonard S. Kisslinger, Int.J.Theor.Phys. D {\bf 16}, 0022R1
  (2016)
\bibitem{tdl19} Yu-Dai Tsai, Patrick DeNeverville and Ming Xiong Liu,
  arXiv:2802590[hep-ph] (2019)
\end{thebibliography}
\end{document}